\documentclass[iop]{emulateapj}
\usepackage{graphicx}
\usepackage{natbib}
\usepackage{footmisc}

\begin{document}

\title{The Relation Between [OIII]/H$\beta$ and Specific Star Formation Rate in Galaxies at $\lowercase{z} \sim 2$}

\author{Claire Mackay Dickey$^{1}$, Pieter G. van Dokkum$^{1}$}
\author{Pascal A. Oesch$^{2}$}
\author{Katherine E. Whitaker$^{3,}$\altaffilmark{$\dagger$}}

\author{Ivelina G. Momcheva$^{4}$}

\author{Erica J. Nelson$^{1}$}
\author{Joel Leja$^{1}$}

\author{Gabriel B. Brammer$^{4}$}
\author{Marijn Franx$^{5}$}
\author{Rosalind E. Skelton$^{6}$}

\affil{$^{1}$Department of Astronomy, Yale University, New Haven, CT 06520, USA; claire.dickey@yale.edu}
\affil{$^{2}$Yale Center for Astronomy and Astrophysics, Department of Astronomy, Yale University, New Haven, CT 06520}
\affil{$^{3}$Department of Astronomy, University of Massachusetts, Amherst, MA 01003, USA}
\affil{$^{4}$Space Telescope Science Institute, 3700 San Martin Drive, Baltimore, MD 21218, USA}
\affil{$^{5}$Leiden Observatory, Leiden University, Leiden, The Netherlands}
\affil{$^{6}$ South African Astronomical Observatory, P.O. Box 9, Observatory, 7935, South Africa}

\altaffiltext{$\dagger$}{Hubble Fellow}

\begin{abstract}

Recent surveys have identified a seemingly ubiquitous population of galaxies with elevated [OIII]/H$\beta$ emission line ratios at $z > 1$, though the nature of this phenomenon continues to be debated. The [OIII]/H$\beta$ line ratio is of interest because it is a main component of the standard diagnostic tools used to differentiate between active galactic nuclei (AGN) and star-forming galaxies, as well as the gas-phase metallicity indicators $O_{23}$ and $R_{23}$. Here, we investigate the primary driver of increased [OIII]/H$\beta$ ratios by median-stacking rest-frame optical spectra for a sample of star-forming galaxies in the 3D-HST survey in the redshift range $z\sim1.4-2.2$. Using $N = 4220$ star-forming galaxies, we stack the data in bins of mass and specific star formation rates (sSFR) respectively. After accounting for stellar Balmer absorption, we measure [OIII]$\lambda5007$\AA/H$\beta$ down to $\mathrm{M} \sim 10^{9.2} \ \mathrm{M_\odot}$ and sSFR $\sim 10^{-9.6} \ \mathrm{yr}^{-1}$, more than an order of magnitude lower than previous work at similar redshifts. We find an offset of $0.59\pm0.05$ dex between the median ratios at $z\sim2$ and $z\sim0$ at fixed stellar mass, in agreement with existing studies. However, with respect to sSFR, the $z \sim 2$ stacks all lie within 1$\sigma$ of the median SDSS ratios, with an average offset of only $-0.06\pm 0.05$. We find that the excitation properties of galaxies are tightly correlated with their sSFR at both $z\sim2$ and $z\sim0$, with a relation that appears to be roughly constant over the last 10 Gyr of cosmic time.

\end{abstract}


\section{Introduction}

The optical emission line doublet [OIII]$\lambda\lambda4959,5007$\AA \ and the recombination line H$\beta$ are diagnostic lines that trace the properties of gas in star-forming regions. H$\beta$ is driven primarily by ionizing radiation, while [OIII] is sensitive to the gas-phase metallicity and the ionization parameter. In combination with other lines, these two diagnostics allow us to measure the metallicity, electron temperature, ionization parameter, and hardness of the radiation field in the interstellar medium (ISM) of a star-forming galaxy \citep{osterbrock}. [OIII]/H$\beta$ is used in the Baldwin-Phillips-Terlevich (BPT) diagram \citep{baldwin81} along with [NII]/H$\alpha$ to distinguish between star-forming galaxies, which lie along the HII abundance sequence, and galaxies powered by active galactic nuclei (AGN) which inhabit a distinct region of BPT parameter space characterized by elevated emission line ratios. For star-forming galaxies, their exact position on the HII sequence is dependent on the galaxy's metallicity, ionizing radiation field, and ISM conditions \citep{kewley01}. 

As rest-frame optical emission lines are being studied at higher redshifts, it appears that galaxies at $z > 1$ exhibit different emission line properties than what is commonly seen in the local universe. Increasingly, it appears that extreme emission line behavior, either very large equivalent widths \citep{maseda14,fumagalli12,vanderwel11,atek11} or enhanced emission line ratios \citep{shapley15,steidel14,holden14,kewley13}, are ubiquitous at higher redshifts.

The physical mechanism driving the observed high emission line ratios remain unclear. Possible culprits include increased electron densities \citep{shirazi14}, harder ionizing radiation fields \citep{steidel14,kewley13}, a larger ionization parameter \citep{brinchmann08b,kewley15cosmos}, increased AGN contributions \citep{forster14,trump11}, shocks \citep{kewley13}, or a combination thereof. 

Recently, several studies have suggested that [OIII]/H$\beta$ correlates with specific star formation rate (sSFR) at low and high redshifts. \cite{kewley15} have found a correlation between sSFR, ionization parameter, and [OIII]/H$\beta$ at $z \sim 0.2-0.6$, and \cite{holden14} have shown that star-forming galaxies at $z\sim3.5$ have similarly high sSFR and [OIII]/H$\beta$ ratios as a small subset of the SDSS star-forming galaxy sample. Additionally, \cite{cowie15} found that at fixed H$\beta$ luminosity, strong emission line diagnostics in general are not shifted from SDSS at high redshift. It may be possible that we are not witnessing an evolution of the [OIII]/H$\beta$ ratio with redshift, but rather that we are probing a population of galaxies with high sSFR which are rare in the local universe but common at higher redshifts. 

A limiting factor is that most $z > 1$ surveys do not sample the low-sSFR regime where the majority of SDSS galaxies lie \citep[e.g,][]{juneau14}. In this work, we utilize the 3D-HST dataset \citep{momcheva15, brammer12a,vandokkum11}, stacking hundreds of galaxies by stellar mass and sSFR respectively. For the first time, we study the [OIII]/H$\beta$ ratio with a sample of galaxies  extending down to $H_{\mathrm{F140W}} \sim 24$ mag, with sSFRs that span two orders of magnitude. 

We assume a flat $\Lambda$CDM cosmology with $\Omega_M = 0.3$, $\Omega_\Lambda = 0.7$, and $H_0 = 70 \ \mathrm{km \ s^{-1} \ Mpc^{-1}}$. All magnitudes are given in the AB system.

\section{Data, Sample Selection, and Stacking}
\subsection{The Dataset}

This analysis is based on the 3D-HST grism dataset\footnote{The 3D-HST dataset may be accessed at http://3dhst.research.yale.edu/Data.php}. The 3D-HST program \citep{vandokkum11,brammer12a,momcheva15} is a 248 orbit near-IR spectroscopic survey with the \textit{HST}/WFC3 G141 grism. 3D-HST provides spatially resolved, low-resolution spectra of all objects in four well-studied extragalactic fields (AEGIS, COSMOS, GOODS-S, and UDS) to a $5\sigma$ depth of $H_{\mathrm{F140W}} \sim 24$ mag. Additional grism data for the GOODS-N field comes from program GO-11600 (P.I: B. Weiner). The G141 grism has a wavelength range of $1.10 \mathrm{\mu m} < \lambda < 1.65 \mathrm{\mu m}$, which gives coverage of H$\beta$ and [OIII]$\lambda\lambda4959,5007$ for the redshift range $1.3 < z < 2.3$.

The photometric catalogs include between 23 and 50 bands of photometry for each field, with \textit{HST}/ACS, \textit{Spitzer}/IRAC, and more. A full description of the catalogs and their creation can be found in \cite{skelton14}. Galaxies in the grism data are identified with co-added $\mathrm{J_{125},\ JH_{140},\ and\ H_{160}}$ WFC3 reference images, and every galaxy spectrum is extracted simultaneously to account for contamination between overlapping spectra. The details of the full extraction and reduction pipelines are fully described in \cite{brammer12a} and \cite{momcheva15}. 

Redshifts were determined using a joint fit to the multi-wavelength photometry and the extracted grism spectra \citep{brammer08}. Comprehensive photometric catalogs have enabled the determination of well-constrained redshifts for even those galaxies that appear devoid of emission lines. In this way, we are able to work with a uniquely unbiased sample of high-redshift galaxies.

Stellar masses and other stellar population parameters were calculated with the FAST code \citep{kriek09}, using simple stellar population models from \cite{bc03}, assuming a \cite{chabrier03} initial mass function, and keeping the redshift fixed to the joint redshift measured above. 

Star formation rates were calculated from the UV+IR luminosities as described in \cite{whitaker14}. In brief, the IR luminosity for each galaxy was converted from the \textit{Spitzer}/MIPS 24 $\mu$m flux density with a luminosity-independent template. The final star formation rates were derived from the combination of the UV and IR luminosities. For those objects in the sample without MIPS detections, we used the UV-derived star formation rate alone.

\subsection{Sample Selection}

The complete 3D-HST photometric catalog contains 207,967 objects, of which 98,663 have grism spectra, going down to $H_{\mathrm{F140W}} \sim 26$ mag. For this work, we required that each spectrum cover $\lambda_{rest} = 4510-5350$ \AA, to allow accurate measurements of the continuum around the emission lines of interest. Our initial sample included 6319 objects in the redshift range $1.395 < z < 2.295$. We removed 1378 objects due to badly fit redshifts or invalid stellar masses or star formation rates as a result of bad photometry. To remove AGN from the sample, we used the IRAC color-color cuts of \cite{lacy2007}. These criteria identify 721 AGN in our sample. For the GOODS-N and -S fields, we also searched for possible x-ray counterparts to our galaxies within 1.5'' in the Chandra Deep North and South Surveys. We found 168 galaxies in our sample that were identified as luminous X-ray AGN with a full 0.5-8 keV luminosity cut \citep[L $>3\times10^{42}$ erg/s;][]{xue2011}, all of which had been successfully selected as AGN via IRAC.

The final sample for this work contains $N = 4220$ galaxies, with stellar masses between $10^{9.2}$ and $10^{11.1} \ \mathrm{M_\odot}$ and specific star formation rates between $10^{-9.6}$ and $10^{-7.9} \ \mathrm{yr^{-1}}$. 

\begin{figure}
\epsscale{1.15}
\plotone{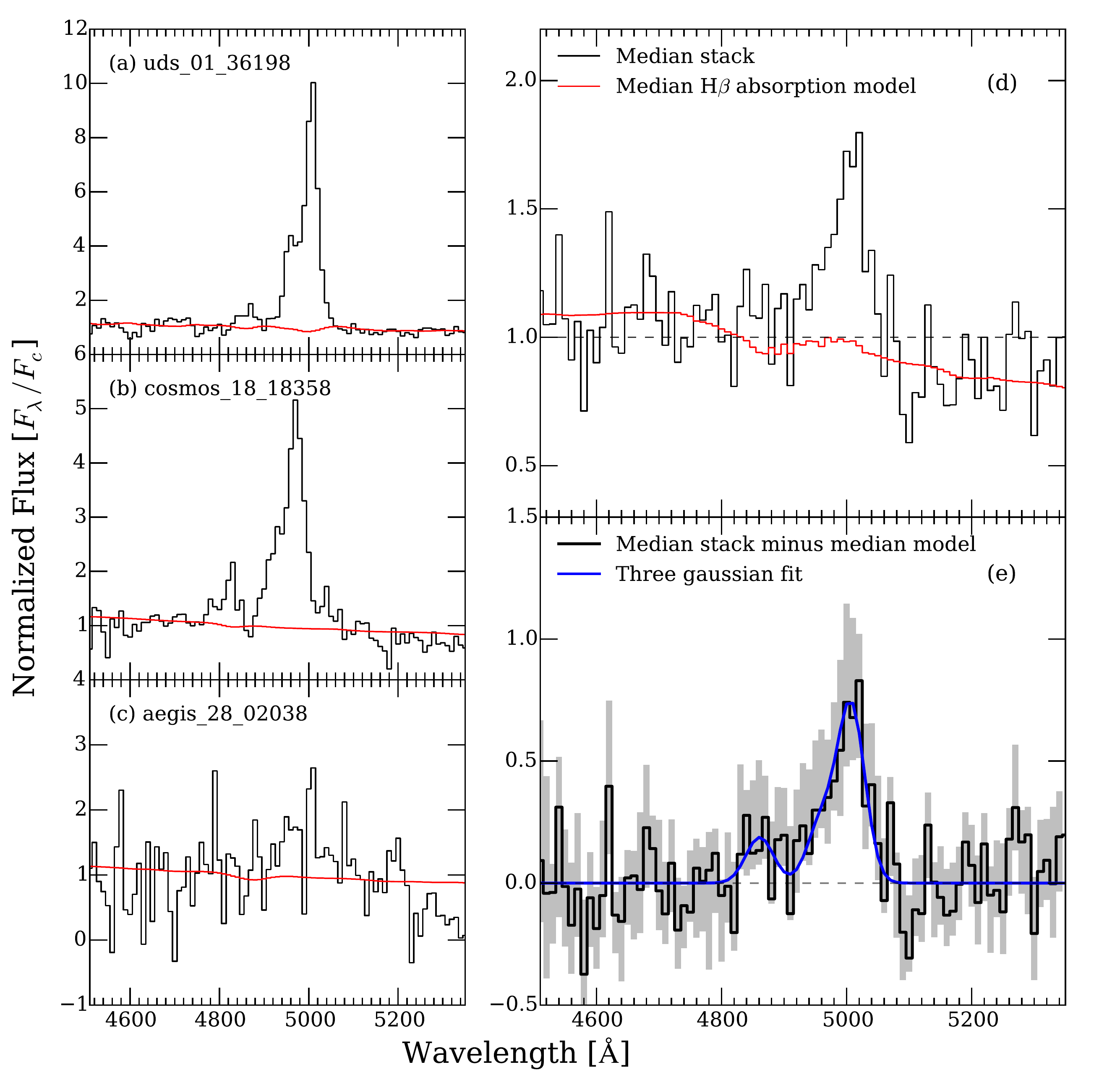}
\caption{Left: three examples of normalized 3D-HST spectra (black) and their corresponding EAZY continuum models (red), all with similar sSFRs. These three objects highlight the diversity of [OIII]/H$\beta$ ratios at $z\sim2$, with some very high ratios (a), some ratios similar to those seen in galaxies at $z\sim0$ (b), and some with very weak emission (c). \\
Right: the median stack of objects with $10^{-8.25} \ \mathrm{yr^{-1}} < \ \mathrm{sSFR} \ < 10^{-8.00} \ \mathrm{yr^{-1}}$, with the median EAZY continuum model in red (d). The continuum-subtracted stack is shown in (e) as the black line, along with a three-component gaussian fit (blue), and the 1$\sigma$ bootstrapped error in grayscale.}
\end{figure}

\subsection{Stacking}

\begin{figure}
\epsscale{1.15}
\plotone{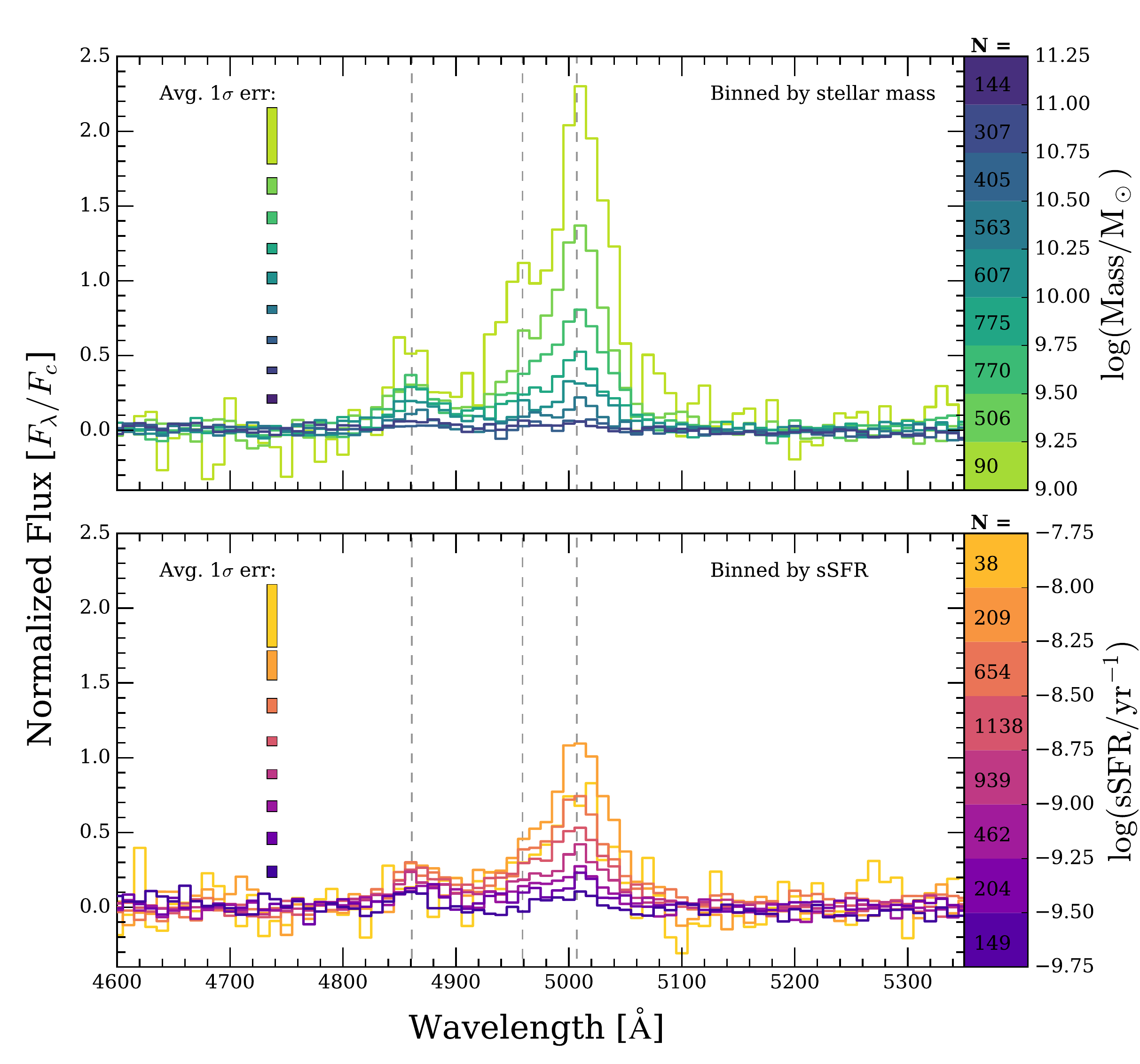}
\caption{Rest-frame 3D-HST spectra, median-stacked in bins of stellar mass (top) and specific star formation rate (bottom ). The shaded regions  represent the average $1 \sigma$ uncertainties at each wavelength, as derived from 100 bootstrapped iterations of the stacking procedure.  Stacking many faint spectra reveals previously undetectable features in the spectra, especially at low sSFRs and high stellar masses. High sSFRs and low masses both produce stronger individual emission lines and larger [OIII]/H$\beta$ ratios.}
\end{figure}

For each object in the sample, we first divide the spectrum and the continuum model by the response function of the grism and shift to the rest-frame. Both the spectrum and the model are projected to a 10-\AA-width rest-frame wavelength grid. In Figure 1(a-c) we show three spectra from our final 3D-HST sample, all from the same sSFR bin. These spectra highlight the variation in emission line properties in our sample. The galaxies shown in Figure 1(a) and 1(b) have clearly detected [OIII] and H$\beta$ emission but different line ratios, with $\log([\mathrm{OIII]/H}\beta) = 1.24\pm 0.03$ and $0.57\pm 0.21$ respectively. 

We median stack the spectra in bins of stellar mass and sSFR for two reasons: first, to utilize the complete dataset, including the objects for which we cannot measure individual emission lines, and second, to test a larger range of masses and sSFRs (especially low sSFRs, where the emission lines are expected to be weaker). We use median rather than mean stacks because while the two methods produce similar ratios in all mass and sSFR bins (within the derived errors), the mean stacks are noisier and badly fit by the continuum models.

Both the mass and sSFR bins have a width of 0.25 dex. Before stacking, we downweight the brightest spectra (those with an [OIII] SNR $> 2$) by a factor of $w = 2/\mathrm{SNR}$ such that there is a smooth transition between weighted and unweighted spectra. The same weighting scheme is applied to the continuum models. We have chosen this weighting threshold because the SNR measurements are unreliable below SNR $\sim 2$, causing a mismatch between the continuum models and the stacked spectra. Using a larger weighting threshold causes the brightest spectra to dominate the stacks, especially at low sSFR. The median stack and model for a single sSFR bin are shown in Figure 1(d). 

Finally, the spectra and corresponding models are both normalized by the average continuum (as measured in the wavelength range 4650-4800 and 5100-5200 \AA) and then the model is subtracted from the median spectrum. This is to account for the effects of stellar Balmer absorption, which is especially significant at low sSFRs and high stellar masses. We find that normalizing by the continuum (as opposed to [OIII] flux, for instance) is necessary at the lowest sSFR and highest stellar masses, as these stacks have very low amounts of emission.

The final stacks are shown in Figure 2. Errorbars for each spectrum are shown as bars at the left in Figure 2 and represent the 16th and 84th percentile values at each wavelength derived from 100 iterations of bootstrap resampling in each bin. We use the 16th and 84th percentile values, as opposed to the standard deviation of the bootstrapped sample, because the underlying distribution is not necessarily gaussian. Unless otherwise noted, all stated 1-$\sigma$ errors refer to the 16th and 84th percentiles of the bootstrapped sample.

To measure the flux of each emission line and the [OIII]/H$\beta$ ratio, we fit each stacked spectrum with three gaussians of equal width, centered at the peaks of each emission line. We require that the [OIII] doublet has a peak ratio of 1:3, as predicted by atomic physics. Using a least squares method, we fit this model to each spectrum (see Figure 1(e) for an example) and measure [OIII]$\lambda5007$/H$\beta$ for each stack. The given uncertainties are the 16th and 84th percentile ratios measured from the 100 bootstrap resamplings of the stacking procedure in each bin. The measured [OIII]/H$\beta$ ratios, errors, and the median properties of each stack are given in Table 1. 

\section{[OIII]/H$\beta$ as a Function of Galaxy Properties}
\subsection{Stellar Mass}

\begin{figure}
\epsscale{1.15}
\plotone{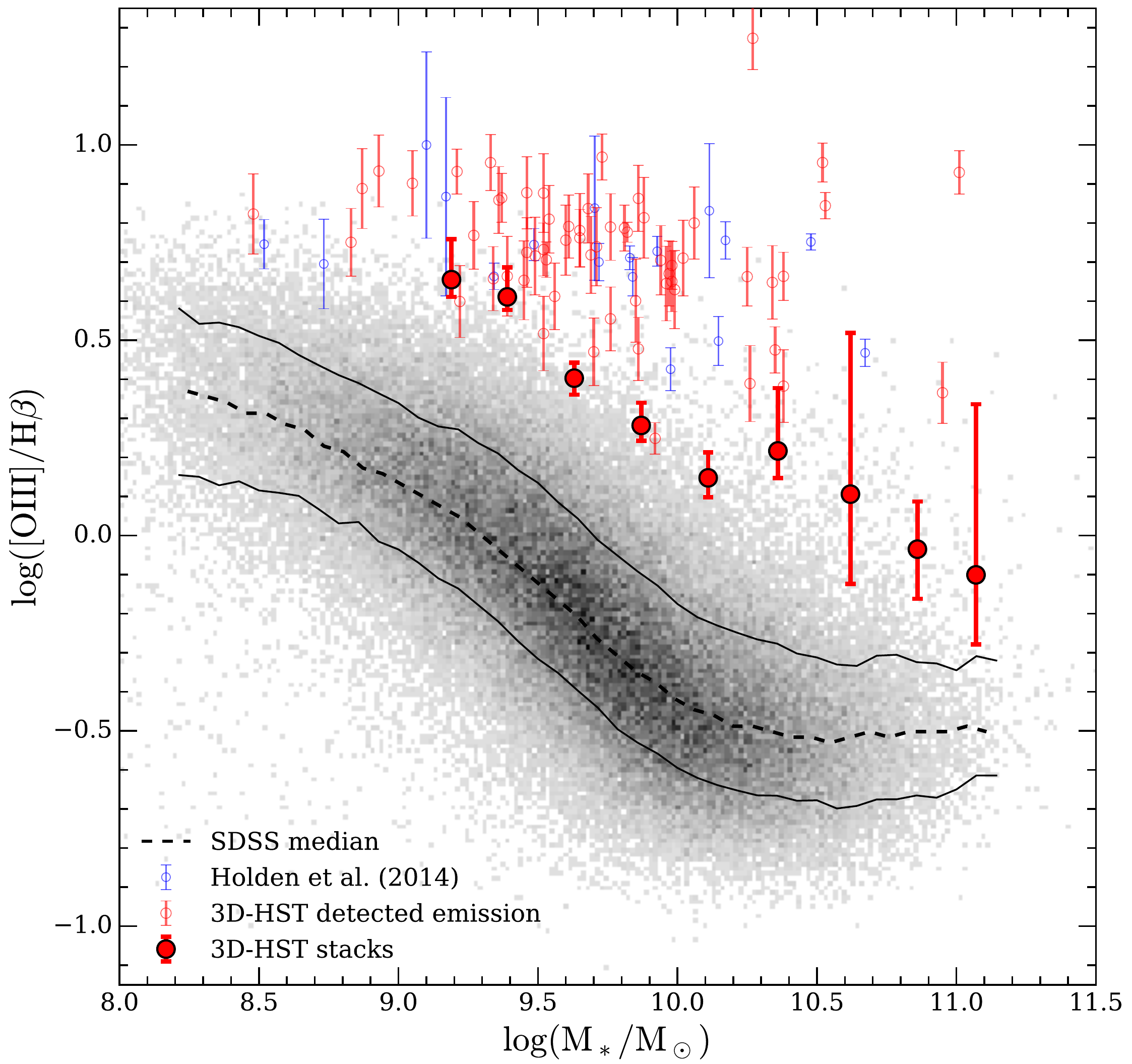}
\caption{The Mass-Excitation (MEx) diagram \citep{juneau11} for the 3D-HST sample (red; z = 1.4-2.3). Individually detected galaxies are shown open-faced, while the median stacks are shown as filled circles. We use the SDSS DR7 star-forming galaxies \citep[with line strengths as measured in][]{brinchmann04} as our $z\sim0$ comparison sample (grayscale), with the median SDSS ratios (dashed) and the 1$\sigma$ contours (solid lines). We also show a sample of individual galaxies at $z\sim3$ from \cite{holden14} (blue). We find that in the 3D-HST sample both individually-detected galaxies and the median stacks have [OIII]/H$\beta$ ratios well above those seen in the local galaxies at fixed stellar mass.}
\end{figure}

\begin{figure*}
\epsscale{1.25}
\plotone{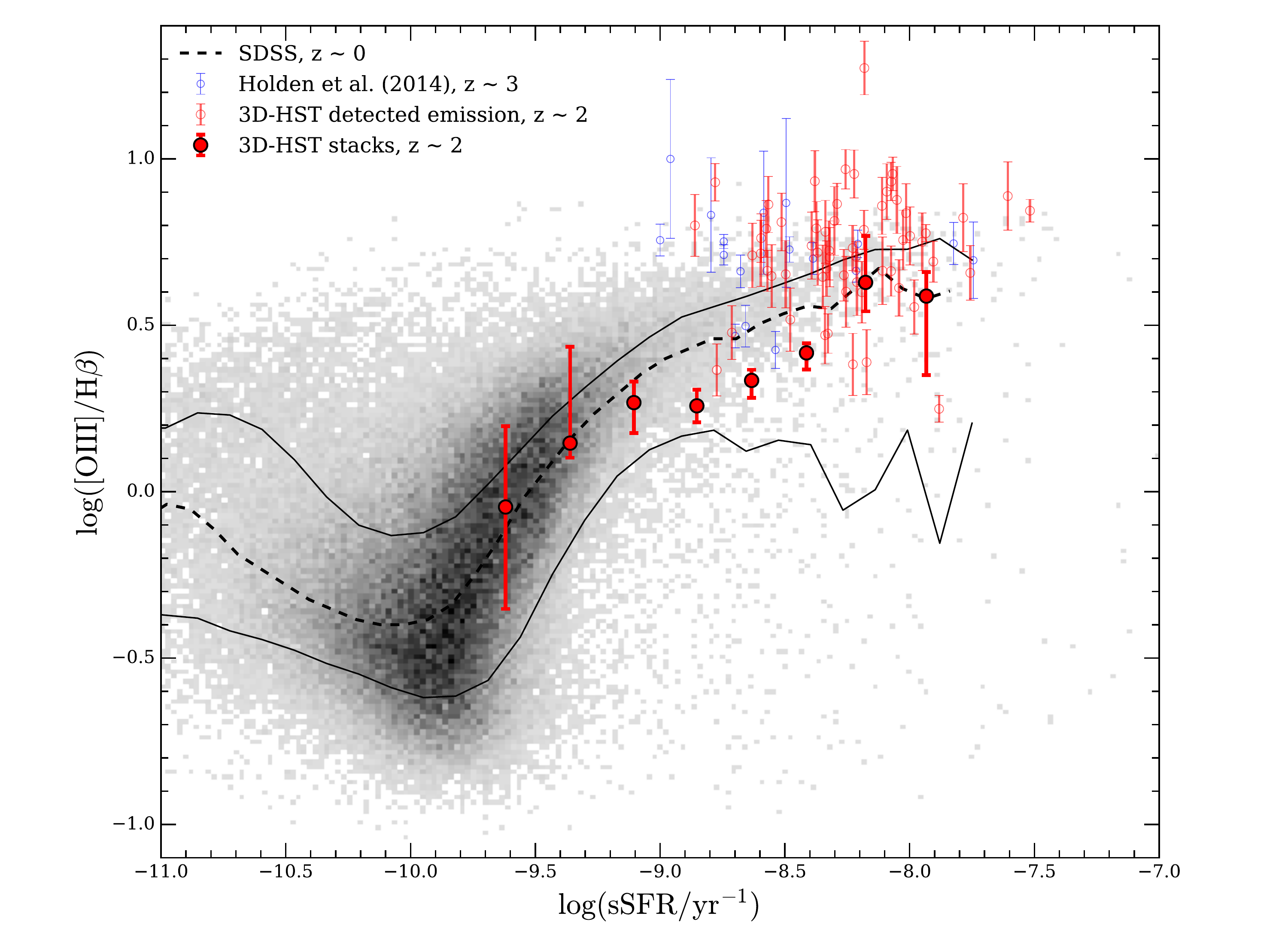}
\caption{The [OIII]/H$\beta$ ratio as a function of specific star formation rate. The 3D-HST data are shown in red, with individual galaxies open-faced and the median stacks as filled circles. Much like the higher redshift individual galaxies from \cite{holden14} (blue), the individually-detected galaxies in 3D-HST lie in the same region of sSFR-excitation space as the SDSS galaxies (grayscale) with the highest sSFRs. More strikingly, the 3D-HST median stacks, which extend to low sSFRs, lie very close the median of SDSS galaxies (dashed) and well within the 1$\sigma$ contours (solid lines). This suggests that the excitation properties of star-forming galaxies are tightly connected to their sSFR, with a relation that is very similar at $z\sim2$ and $z\sim0$.}
\end{figure*}

In Figure 3, we show the [OIII]/H$\beta$ ratio as a function of stellar mass \citep[i.e, the MEx diagram,][]{juneau11}. Our local comparison sample is the star-forming galaxy subset of the SDSS DR7 dataset, classified as such by \cite{brinchmann04} using the \cite{kewley01} BPT diagram demarcations. We also show the median, 16th, and 84th percentile [OIII]/H$\beta$ ratios for the SDSS sample for ease of comparison to the 3D-HST median stacks.

In agreement with previous studies, we find that individual galaxies in the 3D-HST sample have [OIII]/H$\beta$ ratios that are significantly elevated above their SDSS counterparts at all stellar masses. On average, these galaxies have ratios that are $\sim 1$ dex above the median SDSS ratios, in agreement with extant studies \citep[e.g,][shown in Figure 3 in blue]{holden14}.

The median stacks of 3D-HST galaxies binned by stellar mass show similar behavior on a lesser scale. We calculate the average offset between the ratios of the stacked galaxies and the median SDSS values to be $0.59\pm0.05$ dex. This offset, while significant, is much smaller than that observed with the individually detected emission line ratios alone, highlighting the important of stacking galaxies below individual detection limits. Additionally, the stacked galaxies show a strong trend of increasing [OIII]/H$\beta$ ratio with decreasing stellar mass, similar to that seen in the local universe, while the individual galaxies display no such trend.

There is a small jump in the median ratios of the 3D-HST stacks at $\mathrm{M_* \sim 10^{10.25} \ M_\odot}$. While we have removed IRAC-selected AGN from our sample, we acknowledge that these results may still be affected by the presence of undetected AGN and composite galaxies, which may account for the uptick in [OIII]/H$\beta$ at $\mathrm{M_* \sim 10^{10.25} \ M_\odot}$.

\subsection{Specific Star Formation Rate}

In Figure 4, we show [OIII]/H$\beta$ ratios as a function of the specific star formation rate. The SDSS sample shows a tight correlation between sSFR and line ratio above sSFR $\sim 10^{-10} \ \mathrm{yr^{-1}}$. While the majority of high-redshift galaxies with individual detections do not lie within the 16th-84th percentile bounds of the SDSS ratios, there is a subset at sSFR $\sim 10^{-8.75} \ \mathrm{yr^{-1}}$ which do. This is in sharp contrast to Figure 3, where at fixed stellar mass, no $z\sim2$ galaxy in our sample lies within the same bounds. However, the individual galaxies in our sample span a very small amount of sSFR space. By median stacking, we are able to measure line ratios at sSFRs a full dex below those of the individual galaxies.

The line ratios of the median stacks binned by sSFR are strongly correlated with their sSFRs. Most notably, the offset observed between ratios at $z \sim 2$ and $z \sim 0$ at fixed stellar mass is absent in sSFR-space. We measure the average offset between the median ratios of the two populations to be $-0.06\pm0.05$ dex. The strong sSFR - [OIII]/H$\beta$ correlation and the apparent lack of evolution in the relation between sSFR and [OIII]/H$\beta$ across 10 Gyr of cosmic time suggest that the sSFR of a star-forming galaxy is intimately connected to its excitation properties, and that this connection does not change significantly between $z \sim 0$ and $z\sim2$.

\section{Conclusions}

We have investigated the dependence of the [OIII]/H$\beta$ ratio on stellar mass and specific star formation (sSFR) at $z\sim2$, using stacked grism spectra from the 3D-HST survey. By median-stacking many galaxy spectra, most of which have no detected emission on an individual basis, we are able to measure [OIII]/H$\beta$ ratios at unprecedentedly high stellar masses and low specific sSFRs in the high-redshift universe. We use the SDSS DR7 star-forming galaxies as our local universe comparison sample and we find that the [OIII]/H$\beta$ ratios of individual high-redshift galaxies are offset from the average SDSS ratios by $\sim 1$ dex. Median stacking the 3D-HST sample in bins of stellar mass shows that the elevated line ratios at $z\sim2$ are less extreme but still significant, with an average offset of $0.59 \pm 0.05$ dex. 

Median stacking galaxies in bins of sSFR highlights the power of stacking analyses to probe greater swathes of galaxy properties and to avoid the pitfalls of selecting galaxies via emission lines alone. We have shown that [OIII]/H$\beta$ ratios are strongly correlated with sSFR at $z\sim2$ and do not appear to be offset from the median ratios at $z\sim0$ at fixed sSFR. Our results suggest that sSFR is connected to the excitation properties of star-forming galaxies across cosmic time and that elevated [OIII]/H$\beta$ ratios may be driven by high sSFR at $z\sim2$ and $z\sim0$. In this context, the observed evolution of the stellar mass - [OIII]/H$\beta$ ratio relation with redshift may be a consequence of galaxies today having on average much lower sSFRs than galaxies at the peak of cosmic star formation \citep{madau14,whitaker14}.

It has been previously shown that large sSFRs lead to increased ionization parameters and electron densities \citep[e.g.,][]{sanders16,shimakawa15,kewley15cosmos,shirazi14,steidel14}, which in turn can produce the observed large [OIII]/H$\beta$ ratios. There exists a population of local galaxies with similarly ``extreme'' star-formation conditions as to what has been inferred for individual galaxies with large [OIII]/H$\beta$ ratios at high redshift \citep{bian16}. However, for the first time, we show that there also exists a population of galaxies at $z\sim2$ which are implied (through their low sSFRs and low line ratios) to have conditions similar to more typical local star-forming galaxies. 

The exact connection between the physical parameters that govern the [OIII]/H$\beta$ emission line ratio and sSFR remains unclear. Our data, while able to probe a wide swath of the galaxy population, cannot provide insight into the true conditions of star formation in these galaxies, as we cannot measure any of the additional emission lines (e.g, [OII]$\lambda$3727,  [OIII]$\lambda$4363, [NII]/H$\alpha$) which would break the degeneracies inherent in the [OIII]/H$\beta$ measurement. Additionally, contamination from undetected AGN and composite galaxies remains a concern.

Our work highlights the need for spectroscopic surveys that probe the full extent of the galaxy population at high redshift. As has been previously shown by \cite{juneau14} and \cite{cowie15}, sample selection on the basis of luminosity leads to the erroneous conclusion that all galaxies at high redshift exhibit elevated [OIII]/H$\beta$ and high sSFRs. The next generation of 30-meter class telescopes will open up previously inaccessible galaxy populations for spectroscopic surveys, which will be key to understanding the nature of star formation across cosmic time.

\begin{deluxetable}{llll}  
\tablecolumns{4}
\tablecaption{[OIII]/H$\beta$ line ratios from stacked 3D-HST spectra \\ The stacks are binned by the bolded quantity.}

\tablehead{   
  \colhead{N} &
  \colhead{Median sSFR} &
  \colhead{Median $\mathrm{M_*}$} &
  \colhead{[OIII]/H$\beta$ ratio} \\
    \colhead{} &
  \colhead{log($\mathrm{yr}^{-1}$)} &
  \colhead{log($\mathrm{M_\odot}$)} &
  \colhead{}
}
\startdata
\cutinhead{sSFR stacks}\vspace{0.1cm}
38     &  \textbf{-7.93} &  9.57   & $3.87 \pm_{2.11}^{0.65}$ \\  \vspace{0.1cm}
209     &  \textbf{-8.18} &  9.63     & $4.26 \pm_{0.85}^{1.36}$\\ \vspace{0.1cm}
654     &  \textbf{-8.41} &  9.84 & $2.61 \pm_{0.30}^{0.18}$\\ \vspace{0.1cm}
1138     &  \textbf{-8.63} & 9.88	& $2.16 \pm_{0.26}^{0.16}$\\ \vspace{0.1cm}
939     &  \textbf{-8.85} & 9.91   & $1.81 \pm_{0.21}^{0.20}$\\ \vspace{0.1cm}
462     &  \textbf{-9.11} &  10.02  & $1.85 \pm_{0.39}^{0.27}$\\ \vspace{0.1cm}
204 	& \textbf{-9.36} &  10.32 & $1.40 \pm_{0.14}^{0.93}$\\ \vspace{0.1cm}
149     & \textbf{-9.62} & 10.40 & $0.90 \pm_{0.64}^{0.50}$ \\

\cutinhead{Stellar mass stacks}\vspace{0.1cm}
144     & -8.40 	& \textbf{9.19} & $4.52 \pm_{1.07}^{0.46}$ \\ \vspace{0.1cm}
307 	  &  -8.61 	& \textbf{9.39}   & $4.08 \pm_{0.71}^{0.32}$\\ \vspace{0.1cm}
405     &  -8.69 	& \textbf{9.63}    & $2.53 \pm_{0.23}^{0.25}$\\ \vspace{0.1cm}
563     &  -8.72	& \textbf{9.87}     & $1.91 \pm_{0.26}^{0.17}$\\ \vspace{0.1cm}
607     &  -8.74	& \textbf{10.11} & $1.41 \pm_{0.21}^{0.16}$\\ \vspace{0.1cm}
775     &  -8.81	& \textbf{10.36} & $1.65 \pm_{0.61}^{0.26}$\\ \vspace{0.1cm}
770     &  -9.00	& \textbf{10.62}       & $1.28 \pm_{1.21}^{0.68}$\\ \vspace{0.1cm}
506     &  -9.22	& \textbf{10.86}    & $0.92 \pm_{0.26}^{0.27}$ \\ \vspace{0.1cm}
90       &  -9.28	& \textbf{11.07}     & $0.79 \pm_{0.80}^{0.32}$
\enddata
\end{deluxetable}

We thank the anonymous referee for their thoughtful and constructive feedback and their help improving the manuscript. This work is based on observations taken by the 3D-HST Treasury Program (GO 12177 and 12328) with the NASA/ESA HST, which is operated by the Associations of Universities for Research in Astronomy, Inc., under NASA contract NAS5-26555. CMD thanks B.S.\ for their support and many useful discussions. KEW gratefully acknowledges support by NASA through Hubble Fellowship grant \#HF2-51368.001 awarded by the Space Telescope Science Institute, which is operated by the Association of Universities for Research in Astronomy, Inc., for NASA, under contract NAS 5-26555.


\bibliographystyle{apj}


\end{document}